\documentclass[referee]{raa}  

\usepackage{graphicx,times}             
\usepackage{natbib}
\usepackage{amssymb,amsmath}
\usepackage{xcolor}

\bibpunct{(}{)}{;}{a}{}{,}

\begin{document}

\title{Nonlinear Reconstruction of 21cm Global Signal from 21cm Power Spectrum with Artificial Neural Networks}

   \volnopage{Vol.0 (20xx) No.0, 000--000}      
   \setcounter{page}{1}          

   \author{Hayato Shimabukuro
   }

   \institute{Yunnan University, SWIFAR,No.2 North Green Lake Road, Kunming, Yunnan Province,650500, China; {\it shimabukuro@ynu.edu.cn}\\
   Graduate School of Science, Division of Particle and Astrophysical Science, Nagoya University, Chikusa-Ku, Nagoya, 464-8602, Japan}
\vs\no
   {\small Received~~20xx month day; accepted~~20xx~~month day}

\abstract{In this paper, we propose a novel method using artificial neural networks (ANNs) to reconstruct the global 21cm signal from measurements of the 21cm power spectrum. The 21cm global signal provides crucial information on cosmic evolution from the Dark Ages through cosmic dawn and the Epoch of Reionization (EoR). Single-dish telescopes directly measure the global signal, whereas interferometric experiments primarily measure spatial fluctuations, represented by the 21cm power spectrum. While no direct mathematical relationship exists between these two observables—since they probe fundamentally independent Fourier modes—they are indirectly linked through their common dependence on underlying astrophysical and cosmological parameters. The ANN effectively learns this implicit, model-dependent relationship, enabling it to predict the global signal from the power spectrum. We demonstrate that the ANN accurately recovers the global 21cm signal across a broad redshift range ($z=7.5$–$35$) even under realistic observational noise scenarios corresponding to SKA-1 observations. The reconstruction accuracy depends significantly on the spatial scales (wavenumber $k$) included, with larger-scale modes yielding better results due to their stronger sensitivity to global astrophysical processes. Although the ANN method does not provide a model-independent verification of anomalous observations (e.g., the EDGES absorption trough), it offers a computationally efficient and robust tool to infer the global signal within the context of standard astrophysical and cosmological models.
}

   \authorrunning{H.Shimabukuro}            
   \titlerunning{Reconstructing 21cm global signal }  

   \maketitle

%
%

\section{Introduction}

Following the cosmic ``dark ages'' during which no luminous objects existed, the universe witnessed the formation of the first stars and galaxies in a period known as the ''cosmic dawn " \citep [e.g.][]{2001PhR...349..125B}. The X-ray and ultraviolet (UV) photons emitted by these early luminous objects heated and ionized the neutral hydrogen atoms in the intergalactic medium(IGM) \citep[e.g.][]{fur}, leading to the epoch of reionization (EoR), which persisted until the IGM was fully ionized.

The redshifted 21cm line signal from neutral hydrogen is a promising probe for studying the history of the universe from the Dark Ages through the EoR. This signal arises from the hyperfine transition of neutral hydrogen atoms and can provide direct tomographic images of the spatial distribution of HI gas in the IGM \citep[e.g.][]{1990MNRAS.247..510S,1997ApJ...475..429M,fur, 2012RPPh...75h6901P}. Creating three-dimensional maps of this distribution requires high sensitivity and spatial resolution. As an alternative, current radio interferometer arrays, such as LOFAR (Low Frequency Array) \citep[e.g.][]{2013A&A...556A...2V}, the Murchison Widefield Array (MWA) \citep[e.g.][]{2018PASA...35...33W}, and the Hydrogen Epoch of Reionization Array (HERA) \citep[e.g.][]{2017PASP..129d5001D}, aim to statistically detect the 21cm line signal by measuring its power spectrum. These arrays have already set upper limits on the 21cm line power spectrum \citep[e.g., see Fig. 19 of ][]{2023PASJ...75S...1S}, and future experiments, such as the Square Kilometre Array (SKA) \citep{2013ExA....36..235M, 2015aska.confE...1K}, promise to achieve even higher sensitivity. 

Following the discussion of interferometer arrays and their sensitivity to the 21cm line power spectrum, we now turn to the methods of measuring the global 21cm signal. In contrast, single-dish radio telescopes, such as the Experiment to Detect the Global Epoch of Reionization Signature (EDGES)\citep{2018Natur.555...67B}, the Large-aperture Experiment to Detect the Dark Ages (LEDA) \citep{2018MNRAS.478.4193P}, the Probing Radio Intensity at High-Z from Marion (PRIZM)\citep{2019JAI.....850004P}, and the Shaped Antenna measurement of the background RAdio Spectrum (SARAS)\citep{2018ApJ...858...54S,2021arXiv210401756N}, measure the global 21cm signal, representing the sky-averaged brightness temperature. The EDGES has reported an unexpected deep absorption trough at cosmic dawn ($z\sim17$), which has been challenging to reconcile with standard cosmological and astrophysical models\citep{2018Natur.555...67B}. 

Indeed, the tension between the EDGES signal and theoretical models extends beyond its amplitude and involves the overall shape of the observed absorption feature. This discrepancy has sparked debates around non-standard scenarios involving, for example, interactions with dark matter or additional radio backgrounds \citep[e.g.,][]{2018Natur.555...71B,2019MNRAS.486.1763F}. Moreover, subsequent single-dish experiments such as SARAS3 (Shaped Antenna measurement of the background RAdio Spectrum) have independently tested and strongly contested the EDGES detection, further complicating the interpretation.

Different from ground-based telescopes, which are affected by Earth's radio frequency interference and atmospheric absorption, lunar-based or lunar-orbiting telescopes offer significant advantages for detecting the faint 21cm global signal. The farside of the Moon provides a radio-quiet environment, free from Earth's radio frequency interference, making it an ideal location for such sensitive observations. Additionally, the lack of an atmosphere eliminates signal absorption and scattering, allowing for more precise measurements. Currently, several projects for lunar-based or lunar-orbiting telescopes have been proposed, including FarView\citep{2024AdSpR..74..528P}, LuSEE-night\citep{2023arXiv230110345B}, PRATUSH\citep{2023ExA....56..741S}, Discovering the Sky at the Longest Wavelength (DSL)\citep{2021RSPTA.37990566C} and Large-scale Array for Radio Astronomy on the Farside(LARAF)\citep{2024arXiv240316409C}.

Interferometric measurements and single-dish experiments, therefore, provide complementary probes of the early universe. While single-dish telescopes directly measure the global signal, interferometers are sensitive to spatial fluctuations, providing the power spectrum across multiple spatial scales. Although direct measurement of the global signal by interferometers faces inherent challenges due to their baseline configurations\citep{2013PhRvD..87d3002L,2015ApJ...809...18P,2020MNRAS.499...52M,2023ApJ...945..109Z}, attempts to constrain or reconstruct the global signal from interferometric data indirectly have been explored. Notably, recent studies have demonstrated the potential synergy between these two observational approaches, using astrophysical modeling as a common ground\citep{2018MNRAS.478.2193C,2020MNRAS.491.3108F}. 

Several methods have been proposed to relate measurements of the 21cm power spectrum to cosmological observables like the optical depth to the cosmic microwave background (CMB), $\tau$, through modeling of reionization processes\citep[e.g.][]{2016PhRvD..93d3013L,2016ApJ...821...59F,2021PASP..133d4001B,2022JCAP...10..007S,2023PhRvD.108h3531S}. For example, as $\tau$ essentially integrates ionization history, predicting $\tau$ from power spectrum constraints indirectly implies reconstructing the ionization history from 21cm power spectrum measurements, analogous to reconstructing the global signal.

Meanwhile, recent advances in machine learning (ML), particularly artificial neural networks (ANNs, also referred to as neural networks or neural nets), have expanded the analysis methods in 21cm cosmology. Machine learning techniques have been extensively utilized in diverse contexts, including emulation of power spectra\citep[e.g.][]{2017ApJ...848...23K, 2018MNRAS.475.1213S,2019MNRAS.483.2907J,2020MNRAS.495.4845C,2024MNRAS.527.9833B,2024MNRAS.527.9977S}, global signal modeling\citep[e.g.][]{2021MNRAS.508.2923B,2022ApJ...930...79B}, likelihood emulation\citep[e.g.][]{2022ApJ...926..151Z,2022ApJ...933..236Z,2024JCAP...03..027C}, ionized bubble identification\citep[e.g.][]{2021MNRAS.505.3982B,2024MNRAS.529.3684K}, parameter inference\citep[e.g.][]{2017MNRAS.468.3869S,2019MNRAS.484..282G,2019MNRAS.490..371D,2022MNRAS.509.3852P}, treatment of foreground\citep[e.g.][]{2019MNRAS.485.2628L,2021MNRAS.504.4716G} and recovering 21cm statistics from another statistics\citep[e.g.][]{2021MNRAS.506..357Y,2022RAA....22c5027S}. These studies illustrate that ML techniques efficiently capture complicated nonlinear relationships and significantly reduce computational costs in extracting cosmological information from complex datasets, justifying their integration into 21cm cosmological analyses.

Motivated by these developments, in this paper, we propose a novel ANN-based method to predict the global 21cm signal directly from measurements of the 21cm power spectrum. While the global signal and power spectrum probe fundamentally distinct Fourier modes and thus lack a direct, model-independent mathematical relationship, both observables depend implicitly on the same underlying astrophysical and cosmological parameters. ANNs provide a powerful, non-linear modeling tool capable of efficiently capturing these implicit correlations, enabling predictions of the global signal solely from interferometric data. Additionally, ANNs exhibit robustness to realistic observational noise, making them attractive for practical applications to upcoming experiments such as SKA\citep[e.g.][]{2021MNRAS.508.2923B,2022ApJ...930...79B,2022MNRAS.509.3852P}.

While contemporaneous work by \citep{2024arXiv241004792S} has explored the broad potential of ANNs to classify radio background models and map observables bidirectionally, our own work—developed independently and in parallel—proposes a novel method with a different, specific focus. In this paper, we concentrate on predicting the global 21cm signal directly from measurements of the 21cm power spectrum and conduct a detailed investigation into the practical fidelity of this reconstruction.

Although the current study focuses exclusively on demonstrating the ANN's feasibility and robustness, it remains crucial in future work to conduct a detailed quantitative comparison between the ANN-based approach and traditional statistical methods. Specifically, direct comparisons regarding parameter estimation accuracy, computational efficiency, and robustness under realistic observational conditions will be necessary to justify the integration of ANN methods fully within practical 21cm cosmology analyses.


This paper is structured as follows. Section II describes the theoretical framework and simulations used to generate training datasets. Section III outlines the ANN architecture and training procedure. Section IV presents detailed results of the global signal recovery, highlighting the method’s robustness against observational noise. Section V summarizes the findings and discusses implications and future directions for integrating ANN methods within the broader framework of 21cm cosmology analyses.

\section{Cosmological 21cm signal}
\label{sec:21cm}

The fundamental observable for the 21cm signal is the brightness temperature, which can be expressed as:\citep[e.g.][]{2013ExA....36..235M}

\begin{align}
\delta T_{b}(\nu) &= \frac{T_{{\mathrm{S}}}-T_{\gamma}}{1+z}(1-e^{-\tau_{\nu_{0}}})  \nonumber \\
                  &\quad \sim 27x_{\mathrm{H}}(1+\delta_{m})\left(\frac{H}{dv_{r}/dr+H}\right)\left(1-\frac{T_{\gamma}}{T_{\mathrm{S}}}\right) \nonumber \\
                  &\quad \times \left(\frac{1+z}{10}\frac{0.12}{\Omega_{m}h^{2}}\right)^{1/2}\left(\frac{\Omega_{b}h^{2}}{0.024}\right) [\mathrm{mK}].
\label{eq:brightness}
\end{align}

where $T_{\mathrm{S}}$ and $T_{\gamma}$ represent the spin temperature of the IGM and CMB temperature, respectively.  The optical depth in 21cm rest frame at frequency $\nu_{0}=1.4 {\mathrm{GHz}}$ is denoted by $\tau_{\nu_{0}}$. The neutral fraction of hydrogen atom is given by  $x_{\mathrm{H}}$ and $\delta_{m}(\mathbf{x},z) \equiv \rho/\bar{\rho} -1$ represents matter density fluctuations.  The velocity gradient of the IGM along the line of sight is represented by $dv_{r}/dr$, and $H$ is the Hubble parameter.  All parameters are evaluated at redshift $z = \nu_{0}/\nu - 1$. In all of our simulations, we assume a flat $\Lambda$CDM cosmology with parameters consistent with the \textit{Planck} 2018 results: $\Omega_m = 0.315$, $\Omega_b = 0.049$, $h = 0.674$, $\sigma_8 = 0.811$, and $n_s = 0.965$ \citep{2020A&A...641A...6P}.

For practical purposes, it's often more convenient to analyze 21cm fluctuations in Fourier space. The 21cm fluctuations are typically evaluated using the 21cm line power spectrum, defined as: \citep[e.g.][]{fur}
\begin{equation}
\langle \delta T_b(\mathbf{k}) \delta T_b(\mathbf{k^{\prime}})\rangle
= (2\pi)^3 \delta(\mathbf{k}+\mathbf{k^{\prime}}) P_{21}(\mathbf{k}).
\label{eq:ps_def}
\end{equation}

where the dimensionless 21cm line power spectrum is expressed as $k^{3}P_{21}(k)/2\pi^{2}$. In the study, we employ a semi-numerical simulation approach, similar to that used in \citet{2017MNRAS.472.1915C,2018MNRAS.478.2193C} to calculate the 21cm line power spectrum. For more detailed modeling, please refer to these studies.


\section{Artificial Neural Networks}\label{sec:ANN}

In this section, we describe the construction, training, and evaluation of our artificial neural network (ANN; also referred to as neural network or neural net) used to reconstruct the global 21\,cm signal from power‐spectrum inputs.

\subsection{Dataset Preparation}
We generate 500 semi‐numerical 21\,cm models with the code of Cohen et al.\ \citep{2017MNRAS.472.1915C,2018MNRAS.478.2193C}.  Each model provides the 21\,cm power spectrum at three fixed wavenumbers ($k=0.1,\,0.5,\,1.0\,h\,\mathrm{Mpc}^{-1}$) across 76 binned redshift slices from $z=7.5$ to $35$, together with the corresponding sky‐averaged (global) 21\,cm signal. These particular wavenumbers were chosen based on the sensitivity range expected by upcoming interferometric experiments, such as SKA-1, which will provide robust measurements at scales around these values. Of these, 400 models are used exclusively for training and an independent set of 100 models is held out as a test dataset for final performance evaluation; no separate validation set is employed in this study.

\subsection{Network Architecture}
The architecture of our ANN is briefly described as follows. The input data $x_{j}$ is fed to the $j$-th neuron in the input layer. Each input neuron is connected to the $i$-th neuron in the hidden layer with a weight $w^{(1)}_{ij}$ associated with each connection. The $i$-th neuron in the hidden layer, $s_{i}$, is expressed as a linear combination of all input neurons with their respective weights $w^{(1)}_{ij}$:

\begin{equation}
s_{i}=\sum_{j=1}^{n} w^{(1)}_{ij}x_{j},
\label{eq:hidden}
\end{equation}

where $n$ is the number of input data points. In the hidden layer, the $i$-th neuron is activated by an activation function $\phi$, producing the output $t_{i}=\phi(s_{i})$. We use the ReLU function as the activation function, defined as follows:

\begin{equation}
\phi(x)=\mathrm{max}(0,x)=
\begin{cases}
    x & (x\ge 0) \\
    0 & (x<0)
\end{cases}
\end{equation}

In the output layer, the output vector is obtained by computing linear combinations of the activated neurons in the hidden layer with wights $w^{(L)}_{ij}$ (where $L$ denotes the total number of layers):

\begin{equation}
y_{i}=\sum_{j=1}^{k} w^{(L)}_{ij} t_{j},
\label{eq:output}
\end{equation}

where $k$ is the number of neurons in the hidden layer. Note that the output values are not activated. The goal of training the ANN is to find a set of weights that ensures the output vectors produced by the ANN for a set of input vectors are close to the desired output vectors.  Once the weights are adjusted to minimize the difference using a training sample, the ANN can predict output vectors for new input vectors outside the training sample, such as new observational data. To quantify the accuracy of the ANN's output relative to the desired output for the training data, we define the total cost function as:

\begin{equation}
E=\sum_{n=1}^{N_{\mathrm{train}}}E_{n}=\sum_{n=1}^{N_{\mathrm{train}}}\left[\frac{1}{2}\sum_{i=1}^{m}(y_{i,n}-d_{i,n})^{2}\right],
\label{eq:cost}
\end{equation}

where $N_{\mathrm{train}}$ is the number of training datasets, and $m$ is the number of neurons in the output layer. $y$ and $d$ represent the outputs of the ANN and the desired training output data, respectively. Our objective is to minimize the cost function by finding the optimal set of weights. This is achieved by computing the partial derivatives of $E$ concerning the individual weights $w^{(l)}_{ij}$ and finding the local minimum of $E$ using gradient descent. We optimize weights via backpropagation \citep{1986Natur.323..533R} using the Adam optimizer with a fixed learning rate of $1\times10^{-3}$, batch size 20, and 20\,000 iterations.

\subsection{Experimental Settings}

\begin{figure}
    \centering
    \includegraphics[width=1.0\hsize]{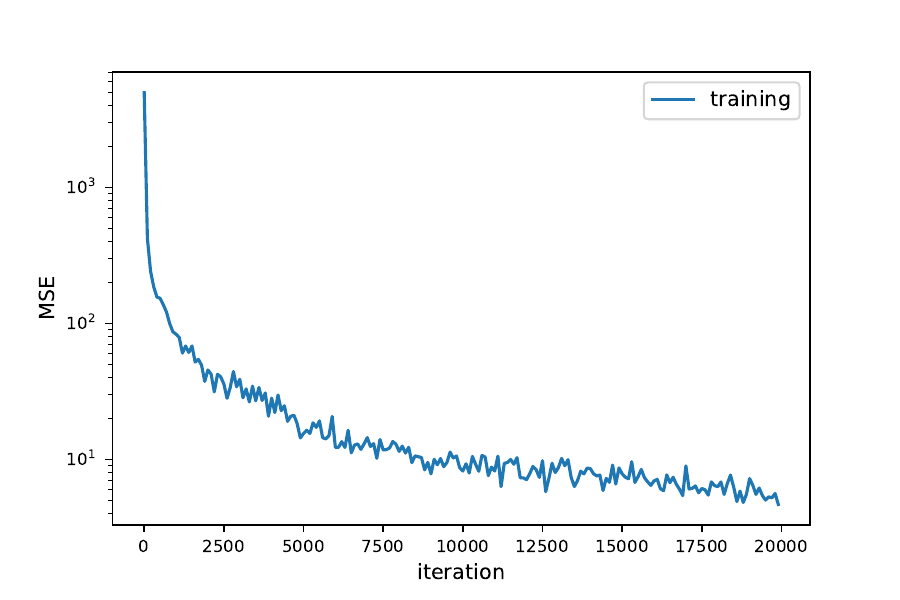}
    \caption{The mean square error (MSE) between obtained and target values in the training dataset. The MSE converges after 20'000 iterations. No validation loss is shown, as no validation set was used in this study. An independent test set of 100 models was held out for final evaluation.
}

    \label{fig:training}
\end{figure}

We performed the backpropagation algorithm with 20,000 iterations for 400 training datasets and then applied the trained network to 100 test datasets. Before discussing the main results, we evaluate whether the training of the ANN architecture is adequate. To do this, we calculated the mean square error (MSE) of the training dataset, defined as

\begin{equation}
\mathrm{MSE}=\frac{1}{N_{\mathrm{train}}}\sum_{i=1}^{N_{\mathrm{train}}}E_{n}
\end{equation}
In Fig. \ref{fig:training}, we show the MSE as a function of the number of iterations. we show the MSE as a function of the number of iterations. We found that the MSE converged after 20,000 iterations. Therefore, we used 20,000 iterations for the backpropagation in subsequent calculations. In this study, no separate validation set was employed: all 400 simulated models were used exclusively for training, and an independent set of 100 models was held out for final performance evaluation as the test set. Accordingly, Fig.~\ref{fig:training} displays only the training loss versus iteration number. The loss curve plateaus by approximately 20\,000 iterations, indicating convergence of the optimization and suggesting that our training dataset is sufficiently large to avoid severe over- or under-fitting. We acknowledge that including a validation loss curve would provide a more complete assessment of the model’s generalization behavior; this will be addressed in future work.

To assess the ANN’s performance under varied conditions, we conduct three classes of experiments, each evaluated on the 100‐model test set.  First, we examine spatial‐scale dependence by training separate networks on single $k$‐bin inputs ($k=0.1$, $0.5$, and $1.0\,h\,\mathrm{Mpc}^{-1}$) to isolate each scale’s contribution to global‐signal recovery.  Second, we test noise robustness by adding SKA‐1 thermal noise and an exaggerated 1000× noise realization to the power‐spectrum inputs.  Finally, we evaluate the ability to recover the high‐redshift cosmic dawn signal by restricting inputs to the EoR‐only range ($7.5\le z\le15$).  In all cases, performance metrics (e.g., correlation coefficient, MSE) are reported on the held‐out test set to quantify generalization.

\section{Results}
In this section, we present the global signal recovered from the 21cm line power spectrum as a function of redshift using the ANN. 

\subsection{Recovered 21cm global signal}
\begin{figure}
    \centering
    \includegraphics[width=1.0\hsize]{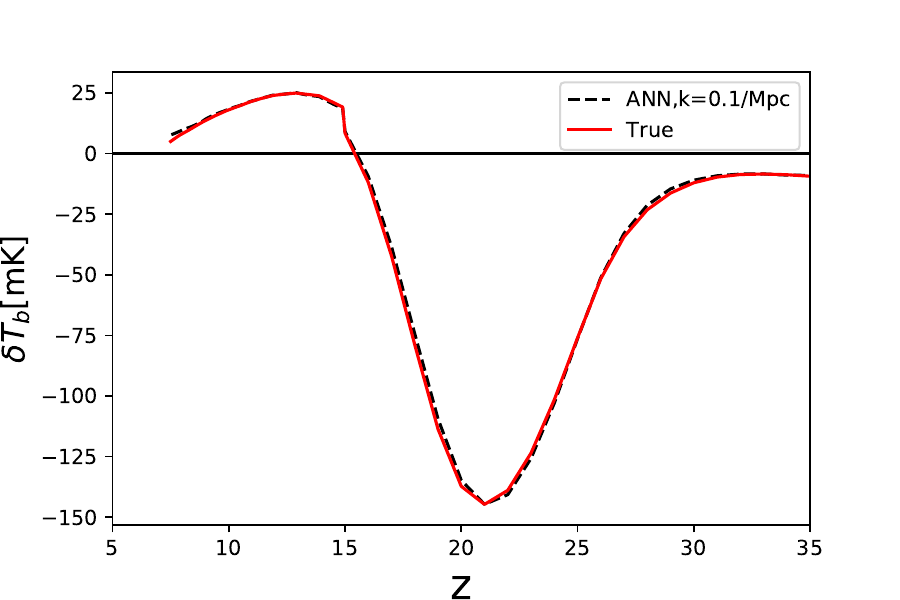}
    \caption{True 21cm global signal({\it red solid}) and recovered 21cm global signal from 21cm line power spectrum from EoR to cosmic dawn($z$=7.5-35)({\it black dashed}).}
    \label{fig:global_signal}
\end{figure}

We recover 21cm global signal at $z=7.5-35$ redshift from the 21cm PS as a function of redshift. We use the 21cm PS at a fixed wavenumber $k=0.1\mathrm {Mpc}^{-1}$. In Fig.\ref{fig:global_signal}, We compare the true global signal with the one recovered using the ANN from the 21cm power spectrum over this redshift range.  As seen in Fig.\ref{fig:global_signal}, the ANN successfully reconstructs the global signal from cosmic dawn to EoR.

Previous studies, such as \citep{2020MNRAS.491.3108F}, have demonstrated that multi-tracer methods can extract the global signal using both 21cm and matter density fluctuations. However, our method, which is based on an ANN, requires only the 21cm fluctuations. The ANN efficiently learns a non-linear mapping between the 21cm power spectrum and the global signal, effectively functioning as a non-linear regression tool that eliminates the need for additional tracers. By training on simulated datasets, the ANN approximates the complex relationship between the 21cm PS input and the global signal, allowing for accurate recovery.

\begin{figure}
    \centering
    \includegraphics[width=1.0\hsize]{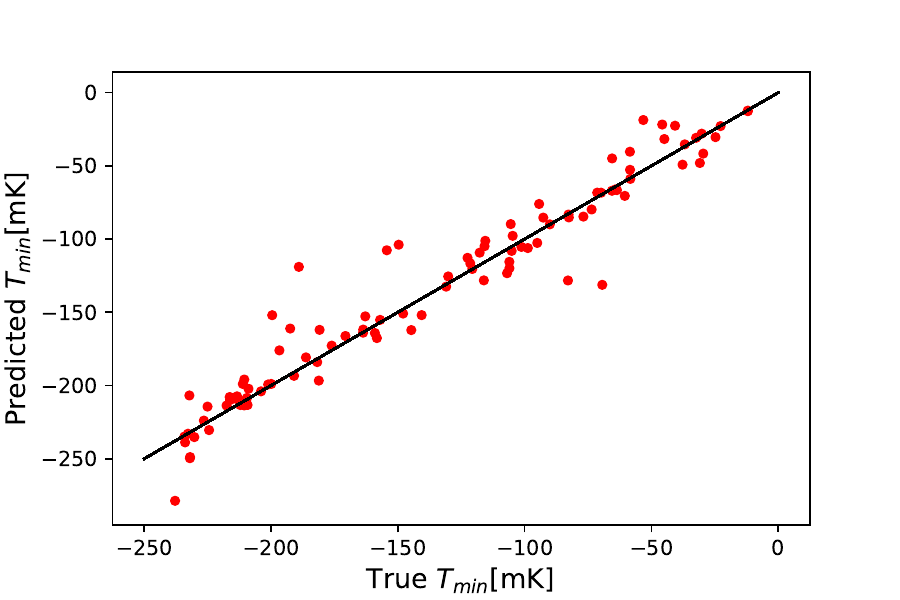}
    \caption{Comparison of depth of trough between obtained and true global signal values for all test datasets. The solid black line (Y=X) represents perfect prediction.}
    \label{fig:T_min}
\end{figure}

To assess the ANN's performance across all test data, we compare the recovered and true global signals at the trough of the 21cm global signal for all test cases. In Fig.\ref{fig:T_min}, we examine the depth of the trough in the recovered global signal across all test datasets, with the Y = X line representing perfect recovery. As shown in the figure, the ANN successfully recovers the depth of the global signal at the trough from the 21cm power spectrum, closely matching the true values. This demonstrates that the model generalizes well across test datasets.

To quantitatively assess the accuracy of the recovered global 21cm signal across all test datasets, we introduce the correlation coefficient (CC), defined as:

\begin{equation}
{\rm CC}=\frac{\sum_{i=1}^{N_{\rm z}}(y_{\rm true,i}- \overline{ y }_{{\rm true}})(y_{\rm ANN,i}- \overline{y}_{{\rm ANN}})}{\sqrt{\sum_{i=1}^{N_{{\rm z}}}(y_{\rm true,i} - \overline{y}_{\rm true})^2}\sqrt{\sum_{i=1}^{N_{\rm z}}(y_{\rm ANN,i}-\overline{y}_{{\rm ANN}})^2}}
\label{eq:cod}
\end{equation}

\noindent where $y_{\rm true, i}$ and $y_{\rm ANN, i}$ represent the true and recovered values of the global signal at redshift $z_i$, respectively, and $N_z$ is the number of redshift bins. The overbars denote mean values averaged over all redshifts. The CC measures the linear correlation between the true and recovered signals for each test dataset; a CC close to 1 indicates a strong positive correlation, while a CC close to $-1$ indicates a strong negative correlation. A higher absolute value of CC signifies a stronger correlation between the datasets.

\begin{figure}
    \centering
    \includegraphics[width=1.0\hsize]{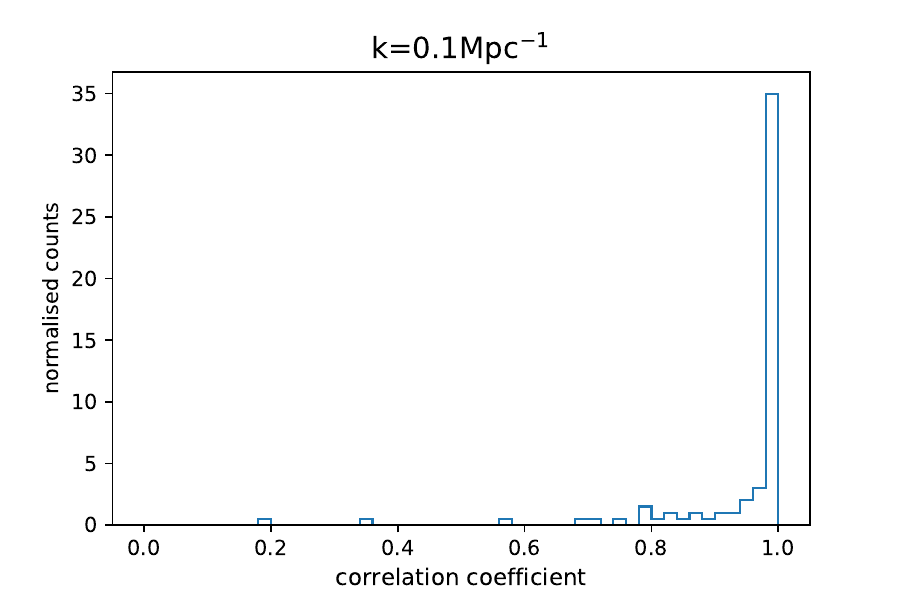}
    \caption{Distribution of the correlation coefficient (CC) for all test datasets where the global signal is recovered from the 21cm PS at $k = 0.1 \mathrm{Mpc}^{-1}$. The mean value is 0.95, and the variance is 0.05.}
    \label{fig:R2_dist}
\end{figure}

We first compute the CC for the case where the 21cm global signal is recovered from the 21cm power spectrum (PS) at $k = 0.1\mathrm{Mpc}^{-1}$. As shown in Fig.\ref{fig:R2_dist}, most of the CC values are distributed between 0.8 and 1.0, with a mean of 0.95 and a variance of 0.05. This quantitatively demonstrates that our artificial neural network (ANN) can successfully recover the 21cm global signal from the 21cm PS at $k = 0.1 \mathrm{Mpc}^{-1}$ for most of the models we consider. However, for some models, the CC values are less than 0.6, indicating that the recovery does not perform as well for these cases compared to others. This can be attributed to the following reasons. In certain models, the 21cm power spectrum at specific scales (i.e. $k=0.1$$\mathrm{Mpc}^{-1}$ does not contain enough information to accurately recover the global signal. This is particularly evident when the power spectrum exhibits anomalies, such as missing the typical three peaks or showing significant shifts in peak positions. Such irregularities hinder the ANN’s ability to effectively learn the relationship between the power spectrum and the global signal. 

\begin{figure}
    \centering
    \includegraphics[width=1.0\hsize]{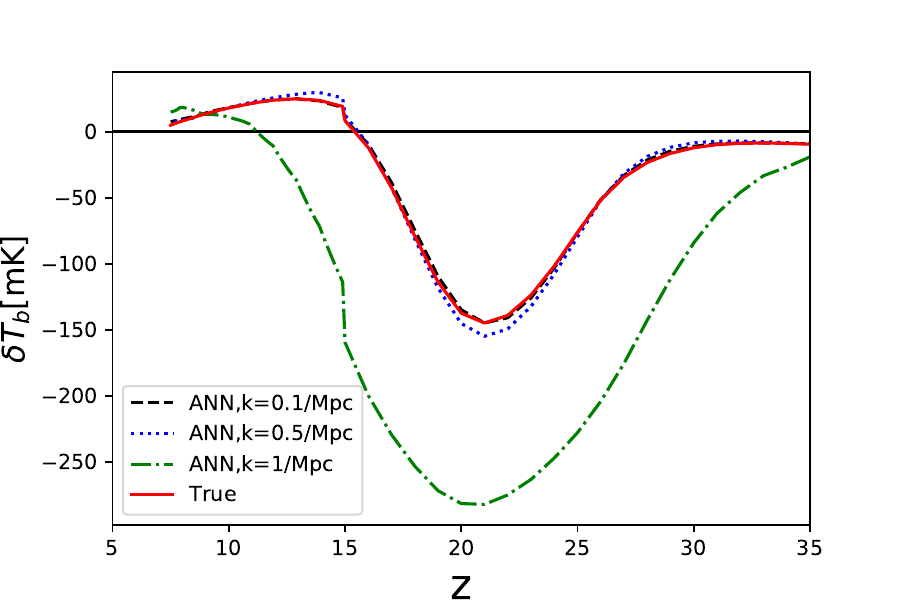}
     \includegraphics[width=1.0\hsize]{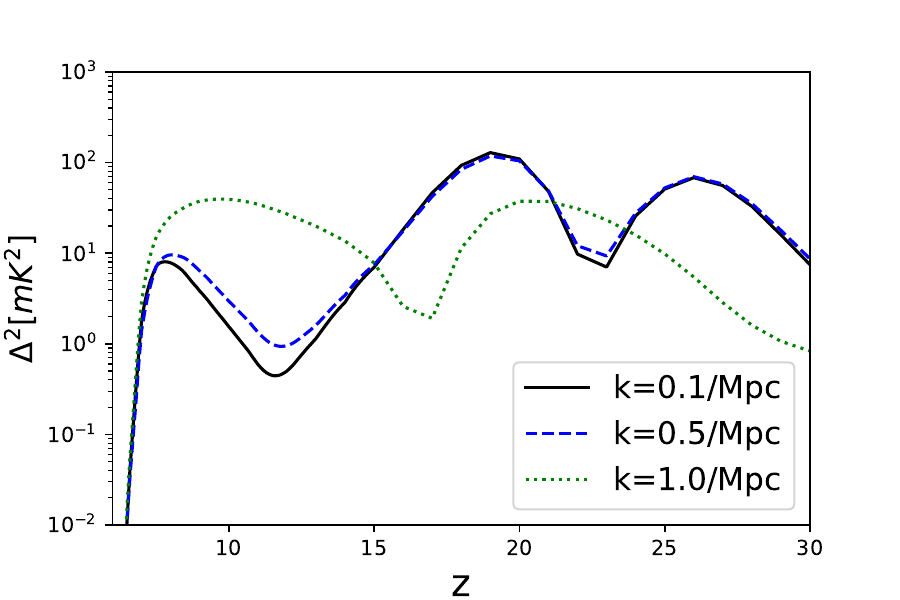}
    \caption{(\textit{Top}) The global 21cm signal recovered using the 21cm PS at fixed wavenumbers $k = 0.1$, $0.5$, and $1.0\mathrm{Mpc}^{-1}$. (\textit{Bottom}) The corresponding 21cm PS as functions of redshift at the same fixed wavenumbers. Note that the recovery deteriorates at $k = 1.0 \mathrm{Mpc}^{-1}$, where the PS exhibits only two peaks instead of three.}
    \label{fig:global_signal2}
\end{figure}

Next, we investigate how the scale of the 21cm power spectrum affects the ANN-based recovery of the global 21cm signal. In Fig.\ref{fig:global_signal2}, we present the recovered global signals and the corresponding 21cm power spectra for fixed wavenumbers $k = 0.1$, $0.5$, and $1.0\,\mathrm{Mpc}^{-1}$. The top panel clearly illustrates that the recovery quality deteriorates at higher wavenumber ($k = 1.0\,\mathrm{Mpc}^{-1}$). The bottom panel highlights the underlying reason: while the power spectra at lower $k$ (0.1 and 0.5~$\mathrm{Mpc}^{-1}$) exhibit three distinct peaks, providing richer and clearer features related to the astrophysical processes across cosmic dawn and reionization, the spectrum at $k=1.0\,\mathrm{Mpc}^{-1}$ lacks these prominent features, thereby offering insufficient information for accurate reconstruction.

It should be noted, however, that the global 21cm signal corresponds mathematically to the spatially averaged brightness temperature and thus represents the $k=0$ Fourier mode. Therefore, the decreased performance at high $k$-values does not result from a mathematical averaging-out of small-scale fluctuations. Rather, this effect arises because larger-scale modes (lower $k$) are more directly sensitive to astrophysical processes that simultaneously shape the global signal, whereas smaller-scale modes (higher $k$) primarily reflect local, small-scale structures whose correlation with the global signal is weaker. Thus, the ANN’s successful recovery of the global signal at lower $k$ reflects its ability to capture indirect, model-dependent astrophysical relationships between these observables rather than a direct mathematical mapping.

\begin{figure}
    \centering
    \includegraphics[width=1.0\hsize]{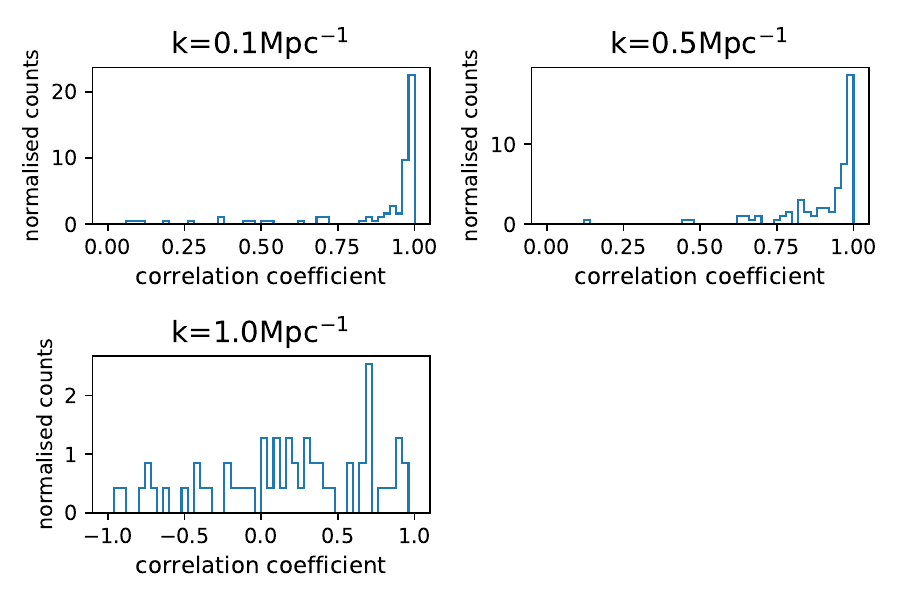}
    \caption{Distribution of the correlation coefficient (CC) for all test datasets where the global signal is recovered from the 21cm PS at different fixed wavenumbers $k = 0.1$, $0.5$, and $1.0\mathrm{Mpc}^{-1}$. The mean CC values are 0.95, 0.88, and 0.18, respectively, with variances of 0.05, 0.18, and 0.27. The recovery accuracy decreases significantly at $k = 1.0\mathrm{Mpc}^{-1}$.}
    \label{fig:R2_dist_2}
\end{figure}

To quantitatively assess the influence of the selected scale on recovery performance, we compute the correlation coefficient (CC) distributions for $k=0.1$, $0.5$, and $1.0\,\mathrm{Mpc}^{-1}$, as shown in Fig.~\ref{fig:R2_dist_2}. For $k=0.1$ and $0.5\,\mathrm{Mpc}^{-1}$, most CC values exceed 0.75, indicating strong reconstruction performance. Conversely, at $k=1.0\,\mathrm{Mpc}^{-1}$, the CC distribution significantly broadens, with a mean value dropping to 0.18 and variance increasing to 0.27. This analysis confirms that larger spatial scales contain more astrophysical information relevant to reconstructing the global signal, reflecting the indirect but meaningful relationship mediated by underlying astrophysical processes.

\subsection{The recovery from the 21cm PS with thermal noise}

So far, we have assumed that the input 21cm line power spectrum is the pure signal derived from simulations. However, in practical observations, measurements of the 21cm line power spectrum are contaminated by random noise. For large radio interferometer arrays like the Square Kilometre Array (SKA), thermal noise dominates the noise budget, although cosmic variance also contributes significantly at large scales. In this subsection, we incorporate both thermal noise and cosmic variance into our analysis to investigate their effects on the reconstruction of the global signal.

The thermal noise power spectrum for a single mode $\mathbf{k}$ is given by \citep{McQuinn:2005hk,Mao:2008ug,2013PhRvD..88h1303M}: 
\begin{equation} 
P_{\mathrm{th},\mathrm{1mode}}(k, \mu) = d_A^2 y \frac{\Omega}{t} \frac{T_{\mathrm{sys}}^2}{\bar{n}({\rm L}{k_\perp}) A_e},
\end{equation} 
where $d_A(z)$ is the comoving angular diameter distance at redshift $z$, and $y(z) \equiv \lambda_{21}(1+z)^2 / H(z)$, with $\lambda_{21} = \lambda(z)/(1+z) = 0.21\mathrm{m}$ and $H(z)$ being the Hubble parameter at $z$. The solid angle of the field of view is $\Omega = \lambda^2 / A_e$, where $\lambda$ is the observing wavelength and $A_e$ is the effective area per station. The total integration time is $t$, and $T_{\mathrm{sys}}$ is the system temperature of the antenna, which is the sum of the receiver temperature (approximately $100\mathrm{K}$) and the sky temperature $T_{\mathrm{sky}} = 60(\nu/300,\mathrm{MHz})^{-2.55}\mathrm{K}$. The term $\bar{n}({\rm L}{k_\perp}) A_e$ represents the number of redundant baselines with L corresponding to $k_\perp$ within a baseline area equal to $A_e$ 

The thermal noise for the mode $\mathbf{k}$ depends on its projection onto the sky plane, $k_\perp = k \sqrt{1 - \mu^2}$, where $\mu = \cos\theta$, and $\theta$ is the angle between the mode $\mathbf{k}$ and the line of sight (LOS).

The thermal noise for the spherically averaged power spectrum over a $k$-shell is given by \citep{2011ApJ...741...70L}:

\begin{equation} P_{\mathrm{thermal}}(k) = \left[ \sum_\mu \frac{N_{\mathrm{c}}(k,\mu)}{P_{\mathrm{th},\mathrm{1mode}}^2(k,\mu)} \right]^{-1/2}, \end{equation} 
where $N_{\mathrm{c}}(k,\mu)$ is the number of modes in the ring with $\mu$ on the spherical $k$-shell with logarithmic step size $\delta k / k = \epsilon$. Specifically, $N_{\mathrm{c}}(k,\mu) = \epsilon k^3 \Delta\mu \times \mathrm{vol} / (4\pi^2)$, and $\mathrm{vol}$ is the survey volume of the sky. The summation accounts for the noise reduction achieved by combining independent modes. It runs over the upper half-shell with positive $\mu$, since the brightness temperature field is real-valued, and only half of the Fourier modes are independent.

The cosmic variance for the 21cm line power spectrum is estimated by 
\begin{equation} P_{\mathrm{cv}}(k) = \frac{1}{\sqrt{N_{\mathrm{modes}}}} P_{21}(k),
\end{equation} where $N_{\mathrm{modes}} = \epsilon k^3 \times \mathrm{vol} / (4\pi^2)$ is the number of modes in the upper half of the $k$-shell.

In this study, we consider an experiment similar to the low-frequency array of SKA Phase~1 (SKA-1). Specifically, we assume a configuration where 224 stations are compactly arranged within a core diameter of 1000 meters, and the minimum baseline between stations is 60 meters. We adopt the following parameters: the field of view of a single primary beam is $\mathrm{FWHM} \sim 3.5^\circ$ at $z \sim 8$, the effective area per station is $A_e \approx 421\mathrm{m}^2$ at $z \sim 8$, the total integration time is 1000 hours, the bandwidth of a redshift bin is 10MHz, and the logarithmic step size of a $k$-bin is $\epsilon = \delta k / k = 0.1$.

Our noise computation results are consistent with previous studies \citep[e.g.,][]{2015aska.confE...1K}. For SKA-1, the cosmic variance is negligible, and the thermal noise is small compared to the signal for $k \leq 1\mathrm{Mpc}^{-1}$. Consequently, the 21cm signal dominates over the noise except at small scales. This favorable signal-to-noise ratio allows for the effective reconstruction of the global signal even in the presence of noise.

\begin{figure}
    \centering
    \includegraphics[width=1.0\hsize]{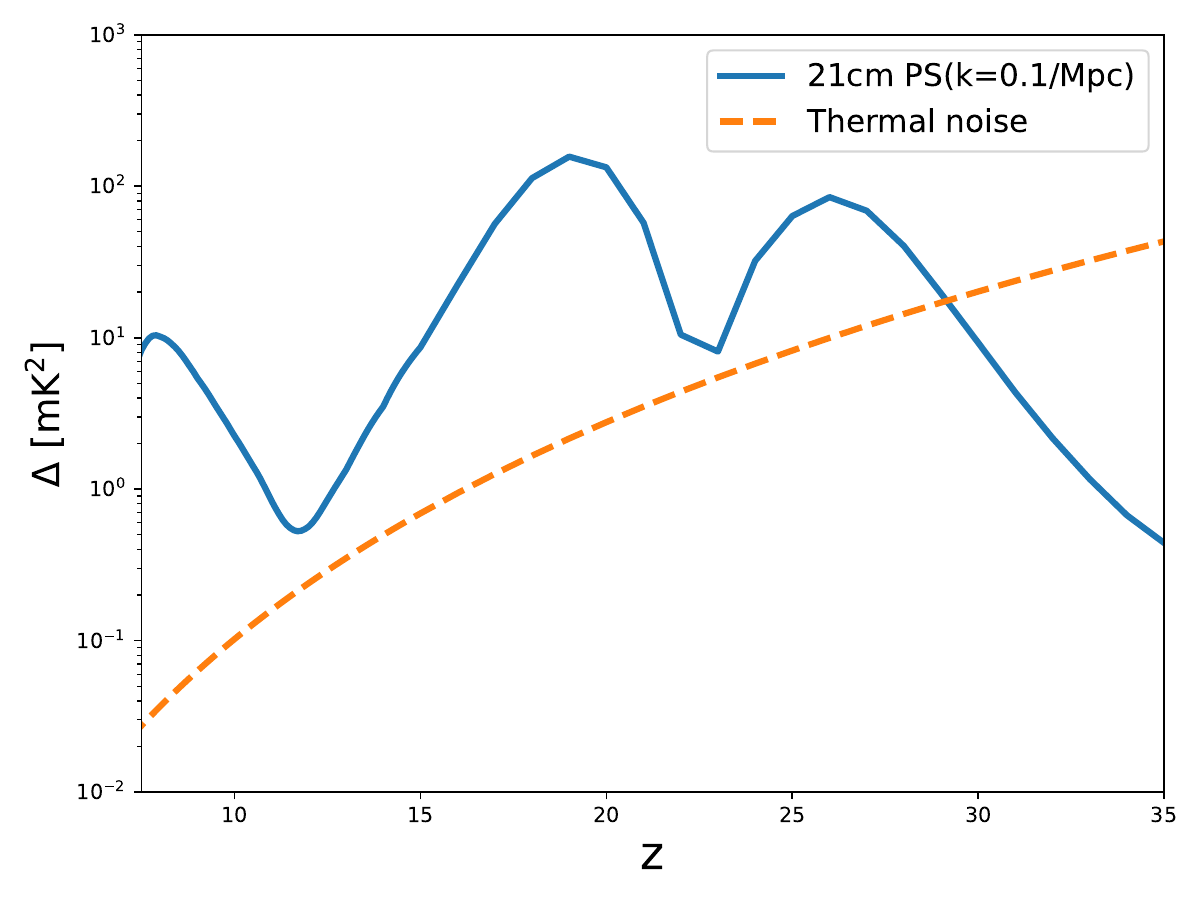}
    \caption{Comparison of 21 cm power spectrum and thermal noise power spectrum. The 21 cm power spectrum (at \(k=0.1\mathrm{Mpc}^{-1}\)) is shown as the solid curve, while the thermal noise power spectrum is shown as the dashed curve.}

    \label{fig:noise_PS}
\end{figure}

We model the measured 21cm line power spectrum as \begin{equation} 
P(k) = P_{21}(k) + N(k), 
\end{equation} 
where $P_{21}(k)$ is the true 21cm line power spectrum signal, and $N(k)$ is a random draw from a Gaussian distribution with zero mean and variance equal to the total noise power spectrum $P_N^2(k) = P_{\mathrm{thermal}}^2(k) + P_{\mathrm{cv}}^2(k)$. In Fig.~\ref{fig:noise_PS}, we show the comparison of the 21 cm power spectrum with the thermal noise power spectrum at various redshifts. At lower redshifts ($z\sim8$), the signal dominates, while at higher redshifts ($z\sim20$), thermal noise becomes comparable to the signal.

\begin{figure}
    \centering
    \includegraphics[width=1.0\hsize]{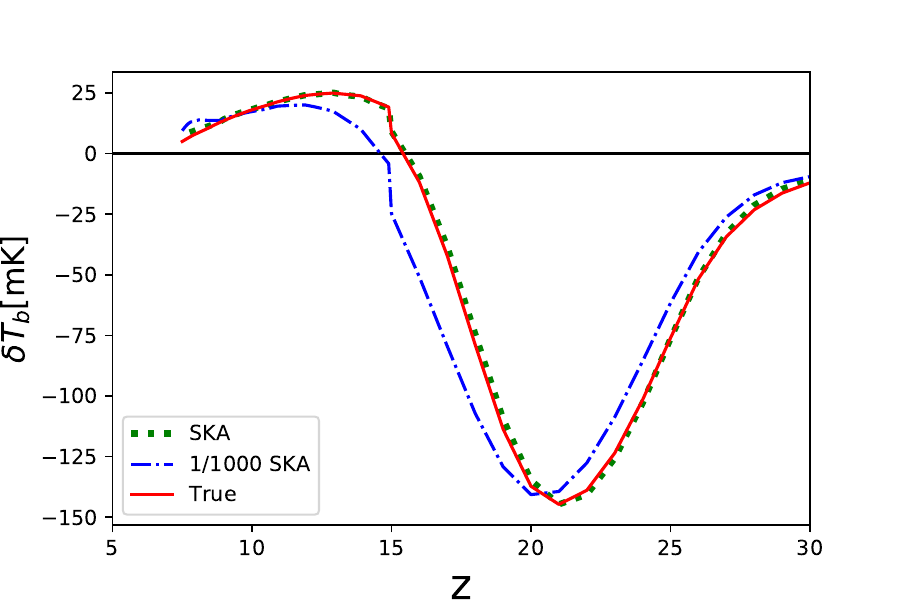}
    \caption{Recovered global signal from noisy 21cm line power spectrum. The target global signal (\textit{red solid line}), recovered signal assuming SKA-1 thermal noise (\textit{green dotted line}), and recovered signal assuming thermal noise 1000 times larger than SKA-1 (\textit{blue dot-dashed line}).}
    \label{fig:gs_noise}
\end{figure}

In Fig.\ref{fig:gs_noise},  we present the global signal recovered from a noisy 21cm line power spectrum assuming SKA-1 experiment specifications. For comparison, we also show the recovered global signal from a 21cm PS with a thermal noise power spectrum that is 1000 times larger than that of SKA-1, which is roughly comparable to the noise levels of MWA or LOFAR \citep{2014MNRAS.439.3262M}. Remarkably, we observe that the 21cm global signal can be adequately recovered from the 21cm PS even when the thermal noise is 1000 times higher than that of SKA-1. This indicates the robustness of our ANN-based recovery method against thermal noise. This result highlights the ANN's ability to effectively mitigate the impact of thermal noise, enabling reliable recovery of the global signal even under challenging observational scenarios. The robustness against such noise levels emphasizes the applicability of this method not only for SKA-1 but also for less sensitive instruments like MWA and LOFAR. 

\begin{figure}
    \centering
    \includegraphics[width=1.0\hsize]{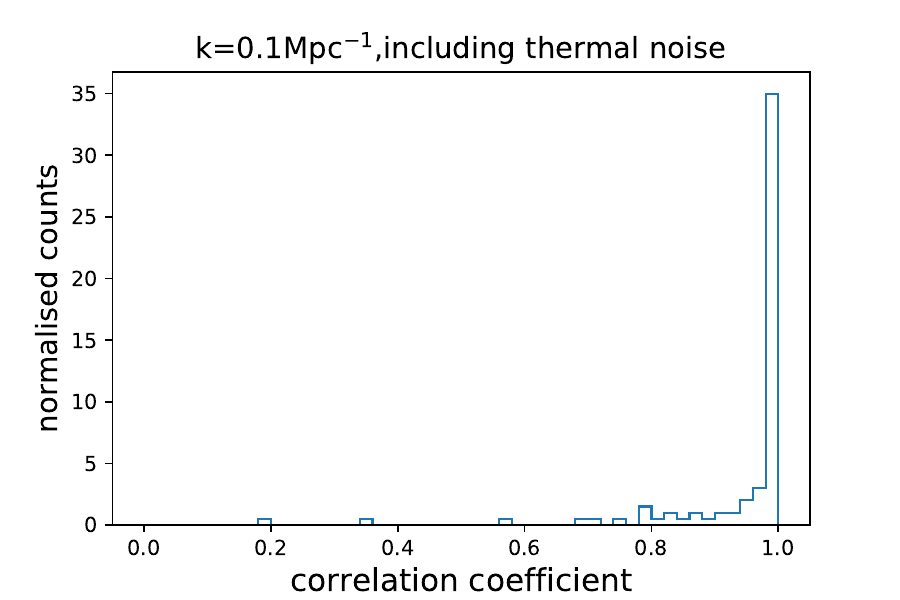}
    \caption{Distribution of the correlation coefficient (CC) for the recovery from the 21cm PS with thermal noise. The mean value and variance of the CC are 0.83 and 0.15, respectively.}
    \label{fig:dist_noise}
\end{figure}

In Fig.\ref{fig:dist_noise}, We display the distribution of the correlation coefficient (CC) for the recovery from the 21cm PS including SKA-1 level thermal noise. Even with thermal noise considered, most of the CC values exceed 0.8, closely resembling the distribution obtained when recovering the global signal from the 21cm PS without thermal noise. This result demonstrates that our ANN maintains high accuracy in reconstructing the global signal despite the presence of thermal noise.

The mean CC value of 0.83 and its variance of 0.15 quantitatively illustrate the resilience of the ANN-based recovery method to observational noise. Such robustness is particularly significant for practical applications, where thermal noise is unavoidable in real observational scenarios. The ability to achieve high recovery accuracy under these conditions emphasizes the suitability of ANN-based approaches for analyzing 21cm data from experiments like SKA-1. Furthermore, this capability supports the potential for cross-validation of recovered signals between interferometric and single-dish observations, enhancing the reliability of 21cm cosmological studies.

These results also underline the ANN's potential to handle complex observational noise environments without significant loss of accuracy. The method's consistent performance across varying noise levels reinforces its role as a robust tool for bridging different observational strategies and extracting meaningful cosmological information from noisy datasets.

To further assess the robustness of our method, we investigate whether the global signal from the cosmic dawn to the EoR can be reconstructed using only the 21cm PS at redshifts corresponding to the EoR. Specifically, we use the 21cm PS (without thermal noise) at redshifts $z = 7.5--15$, corresponding to the EoR, and at multiple wavenumbers ($k = 0.1$--$1.0\mathrm{Mpc}^{-1}$, divided into 30 bins). In this scenario, the input to the ANN consists of the 21cm PS limited to $z = 7.5$--15, while the output layer still covers the broader redshift range  $z = 7.5--35$ for the global signal. The total number of input neurons is 2310 (77 redshift bins $\times$ 30 wavenumber bins), adjusted to match the reduced redshift range of the input PS.

\begin{figure}
    \centering
    \includegraphics[width=1.0\hsize]{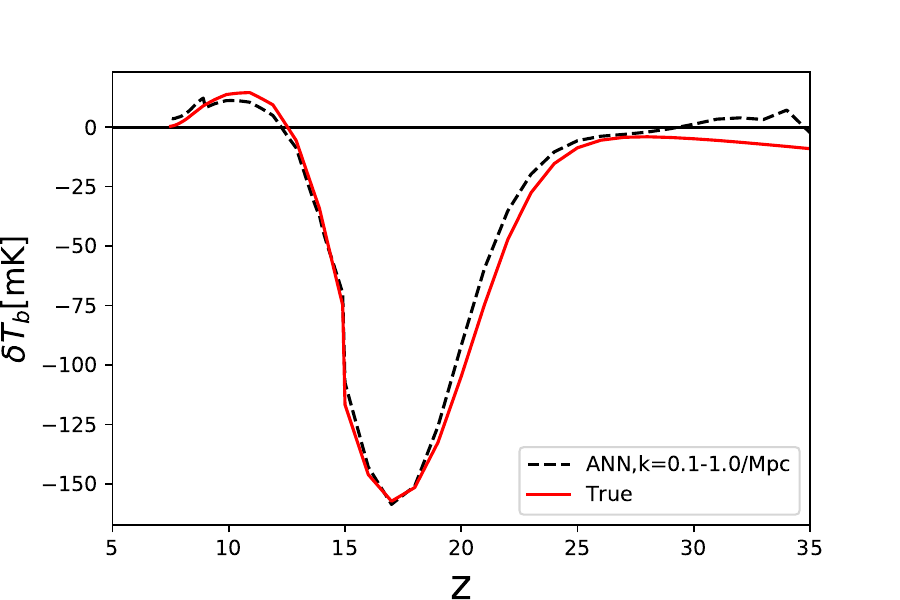}
    \includegraphics[width=1.0\hsize]{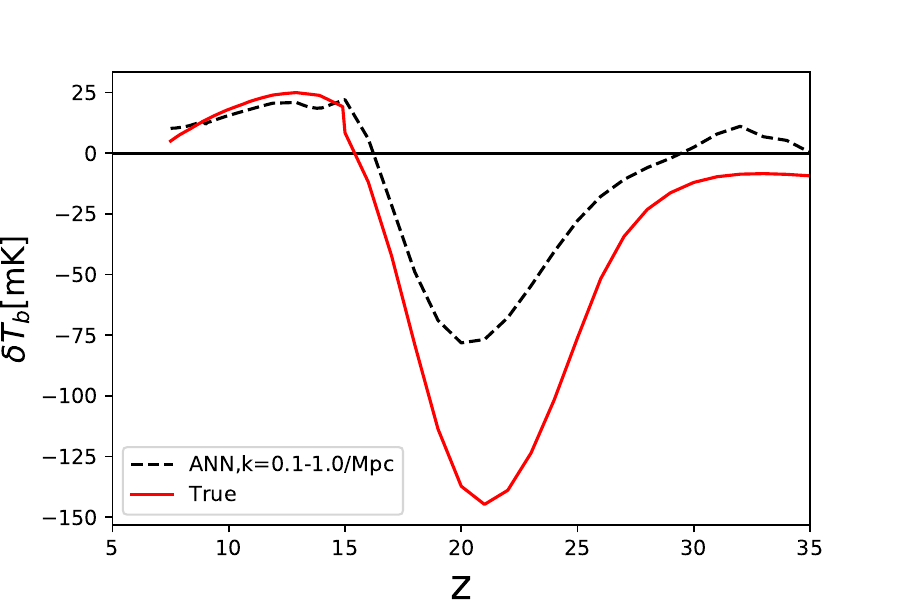}
    \caption{Examples of the recovered 21cm global signal using the 21cm line power spectrum during the EoR at $z = 7.5$--15. The \textit{red solid line} represents the target global signal, while the \textit{black dashed line} shows the recovered signal from the EoR PS.}
    \label{fig:global_signal3}
\end{figure}

In Fig. \ref{fig:global_signal3}, we present examples of the recovered global signal using only the 21cm PS during the EoR. This figure evaluates whether the power spectrum at EoR redshifts contains sufficient information to reconstruct the global signal over a broader redshift span. For one specific model, the ANN successfully reconstructs the global signal across the redshift range, suggesting that certain astrophysical processes during the EoR leave a strong imprint on the power spectrum that correlates with the evolution of the global signal. However, the ANN fails to accurately reconstruct the global signal for another model, indicating that the EoR PS alone does not always encode sufficient information about earlier epochs.

\begin{figure}
    \centering
    \includegraphics[width=1.0\hsize]{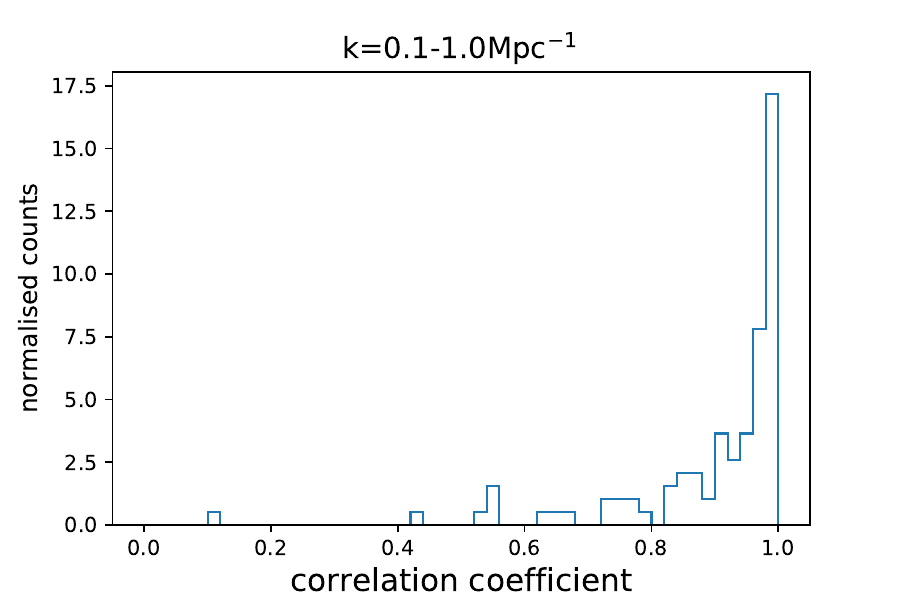}
     \caption{Distribution of the correlation coefficient (CC) for all test datasets when recovering the global signal using the 21cm PS during the EoR. The mean CC is 0.774, and the variance is 0.0677.} 
    \label{fig:R2_2}
\end{figure}

In Fig.\ref{fig:R2_2}, we show the distribution of the correlation coefficient (CC) for all test datasets in this scenario. The distribution reveals a mean CC value of 0.774 with a variance of 0.0677, indicating moderate success overall. However, the wide spread of CC values highlights significant variability among individual models. For certain models, high CC values (close to 1) suggest that the 21cm power spectrum during the EoR contains sufficient information to infer the global signal from cosmic dawn. This is likely due to strong correlations between astrophysical processes during reionization, such as X-ray heating or early star formation, and the thermal and ionization history of the intergalactic medium (IGM) during cosmic dawn. In these cases, large-scale features in the EoR PS, such as ionization bubbles, act as effective proxies for earlier cosmic conditions. Conversely, lower CC values observed for some models indicate that the 21cm PS at the EoR alone does not always encode the necessary information for accurate reconstruction. This discrepancy can arise in scenarios where the processes governing the 21cm PS during reionization are weakly coupled to the thermal evolution of the IGM at earlier epochs. For example, rapid reionization or minimal X-ray heating may reduce the imprint of cosmic dawn on the 21cm PS at the EoR, resulting in less informative features.

\section{Summary \& Discussion}

In this paper, we have introduced a novel method using artificial neural networks (ANNs) to recover the global 21cm signal from the 21cm power spectrum. This approach leverages the ANN’s ability to capture complex, non-linear relationships between these two observables, which, despite lacking a direct model-independent mathematical connection, share dependencies on common astrophysical and cosmological parameters. our results indicate that the ANN-based method can accurately reconstruct the global 21cm signal across a broad redshift range ($z=7.5$–$35$), achieving high correlation coefficients (CC~$>0.8$) even when realistic observational noise levels similar to those anticipated for SKA-1 are included.

The recovery accuracy, however, significantly depends on the spatial scales (wavenumbers $k$) of the power spectrum used as ANN input. Larger scales (lower $k$ modes, e.g., $k=0.1,\,0.5\,\mathrm{Mpc}^{-1}$) typically yield better reconstruction of the global signal, since these scales carry astrophysical information closely correlated with large-scale cosmic processes during cosmic dawn and the epoch of reionization. In contrast, smaller scales (higher $k$), which primarily probe small-scale astrophysical processes, are less effective at constraining the sky-averaged global signal. It is important to clarify that this difference in recovery accuracy arises not from a direct mathematical averaging-out of small-scale fluctuations (as the global signal corresponds strictly to the $k=0$ Fourier mode), but rather from the indirect, model-dependent correlations introduced by the underlying astrophysical parameters.

The ANN-based approach presented here is fundamentally model-dependent, relying entirely on the training data generated from physically motivated astrophysical and cosmological parameters. Consequently, while the method offers a promising indirect way to infer the global signal from interferometric data, it cannot provide a genuinely independent validation of anomalous signals, such as the unexpectedly deep 21cm absorption trough reported by EDGES. Specifically, if the observed global signal reflects physics beyond the scenarios included in the training dataset, the ANN approach would not identify it as novel physics but rather interpret it within the limited parameter space it has learned. Hence, future work must focus on expanding the training dataset to include broader parameter ranges and non-standard physical scenarios, enhancing the ANN’s ability to detect or distinguish unexpected physical phenomena.

Additionally, the robustness of this ANN approach hinges upon the comprehensiveness and representativeness of the training datasets. While the current study varied astrophysical parameters within plausible ranges, it is essential for future studies to systematically explore the sensitivity of ANN predictions to the choice of training parameters and model assumptions. Furthermore, the uncertainty quantification of ANN predictions remains a significant challenge. Incorporating Bayesian neural network techniques or ensemble-based uncertainty estimations will be crucial in future extensions to quantify the confidence in reconstructed signals properly.

In this study, we focused on the ability of our ANN to reconstruct the global 21 cm signal in the presence of thermal noise. However, for upcoming experiments such as HERA, foreground contamination, particularly from galactic synchrotron radiation, will be a significant challenge.  HERA is expected to provide higher sensitivity measurements of the 21 cm signal, but foregrounds, especially at large angular scales, may dominate the observed signal.  While our ANN has demonstrated robustness under thermal noise conditions, its ability to accurately reconstruct the 21 cm signal in the presence of foregrounds remains an open question. Future work will need to explore how well our method can generalize to these foreground-dominated regions and whether additional techniques such as foreground subtraction or model-based correction can be integrated with the ANN to enhance its performance. We suggest that a more comprehensive study of the combination of ANN and foreground removal methods will be essential for improving the accuracy of signal reconstruction in upcoming HERA observations.

In conclusion, the ANN-based method developed in this study offers a valuable complementary analytical tool for 21cm cosmology, particularly useful due to its computational efficiency and capability to capture complex nonlinear relationships. While acknowledging its inherent model-dependence and current limitations, this approach represents a meaningful advancement in utilizing machine learning techniques to extract astrophysical information from upcoming 21cm observations.

\section*{Acknowledgements}

We appreciate Yi Mao's for useful comments and Anastasia Fialkov for providing simulation datasets. This work is supported by the National SKA Program of China (No.2020SKA0110401) and NSFC (Grant No.~12103044).

\bibliographystyle{raa}
\bibliography{reference}

\end{document}